\documentclass[12pt]{article}

\usepackage[margin=1in]{geometry}
\usepackage{graphicx}
\usepackage{caption}

\usepackage{amsmath}
\usepackage{amssymb}
\usepackage{booktabs}
\usepackage{tikz}
\usepackage{verbatim}
\usetikzlibrary{arrows}

\usepackage{subcaption}
\usepackage{soul}

 \newtheorem{theorem}{\bf Theorem}[section]
 \newtheorem{lemma}{\bf Lemma}[section]

 \newtheorem{@definition}{\bf Definition}[section]
 \newenvironment{definition}{\begin{@definition}\rm}{\end{@definition}}
 
 \newtheorem{@example}{\bf Example}[section]

\newtheorem{@nonexample}{\bf (Non)Example}[section]

 \newtheorem{@remark}{\bf Remark}[section]

\begin{document}

\begin{center}
{\Large \bf Tsirelson Polytopes and Randomness Generation}

\medskip

Peter Bierhorst\footnote{Mathematics Department, University of New Orleans, Louisiana, USA} and Yanbao Zhang\footnote{NTT Basic Research Laboratories and NTT Research Center for Theoretical Quantum Physics, Japan}
\end{center}

\begin{abstract}
We classify the extreme points of a polytope of probability distributions in the (2,2,2) CHSH-Bell setting that is induced by a single Tsirelson bound. We do the same for a class of polytopes obtained from a parametrized family of multiple Tsirelson bounds interacting non-trivially. Such constructions can be applied to device-independent random number generation using the method of probability estimation factors \cite{zhang:2018}. We demonstrate a meaningful improvement in certified randomness applying the new polytopes characterized here. 
\end{abstract}

\section{Introduction}

The phenomenon of Bell nonlocality \cite{BELL} is a prediction of quantum physics in which entangled particles display behavior incompatible with any local realistic (or ``classical'') explanation. Originally discovered during an examination of foundational assumptions about physics, Bell nonlocality was later found to have applications to tasks in quantum information theory. Specifically, communication protocols based on an underlying Bell nonlocality experiment can be designed to be ``device-independent,'' in the sense that users of the protocols can be assured of security so long as the observed data meets certain statistical benchmarks, without having to make detailed assumptions about the internal functioning of their devices. Applications include device-independent quantum key distribution \cite{barrett:2005,acin:2007} and device-independent random number generation \cite{colbeck:2007,colbeck:2011,pironio:2010,bierhorst:2018,liu:2018,zhang:2020}.

A Bell experiment will generate measurement outcomes according to a probability distribution. Only certain probability distributions exhibit Bell nonlocality while others do not, and only those exhibiting Bell nonlocality are resources for device-independent tasks. The set of quantum-achievable probability distributions that do exhibit Bell nonlocality is hard to characterize with a curved boundary \cite{navascues:2007,branciard:2011}. Even in the simplest (2,2,2) Bell scenario (2 parties, 2 settings, 2 outcomes), new discoveries about the structure of this set are still being made \cite{goh:2018}.

Here, we describe a method for approximating the set of quantum-achievable distributions from the outside with convex polytopes,
motivated by an application to device-independent random number generation \cite{zhang:2018,PEF}. We obtain the approximation by first restricting to the set of no-signaling probability distributions -- itself a polytope containing the quantum set \cite{barrett:2005} -- and then reducing to a smaller polytope by intersecting with half-spaces defined by so-called ``Tsirelson'' inequalities: linear constraints obeyed by quantum-achievable probability distributions. We did this in an earlier work \cite{PEF} for single instances of Tsirelson's original inequality \cite{tsirelson:1993}; in the current paper, we generalize and formalize this approach to apply to any generalized Tsirelson inequality such as those found in Refs.~\cite{acin:2012,wolfe:2012}. We also simultaneously incorporate multiple Tsirelson inequalities that interact non-trivially and describe the resulting smaller polytopes. While the extreme points of a given polytope characterized by known linear constraints can be found algorithmically, our work goes beyond this in classifying polytopes as parametrized families depending on parameters of the generalized Tsirelson inequalities that induce them. These analytic results provide a useful tool for optimizing approximating polytopes for a given task.

The specific application motivating our study is the Probability Estimation Factor (PEF) method \cite{zhang:2018,PEF} for device-independent random number generation. As demonstrated recently \cite{zhang:2020,shalm:2019,li:2019} the PEF method is effective for certifying randomness in feasible experiments. In its basic form it certifies randomness secure against an adversary holding classical side information, as done in \cite{shalm:2019}, and it can also be used \cite{zhang:2020,li:2019} as a tool for constructing the necessary machinery to execute the quantum probability estimation protocol of \cite{zhang2:2020} which is secure against more general quantum side information. 

The relevance of bounding polytopes to the PEF method is explained in detail in Section \ref{s:applications} below, but it can be understood roughly as follows: if a candidate for a randomness certifying function is found to be valid, loosely speaking, for a finite set of probability distributions, then it will be valid for any convex mixture of these probability distributions. If the set of these convex mixtures -- which is a polytope -- contains the entire quantum set, the function is then confirmed to be appropriate for certifying randomness, having only had to check the validity condition for a finite number of probability distributions. In contrast to the polytopes presented here, other methods for approximating the quantum set, such as the non-linear methods of Navascu\'es, Pironio, and Ac\'in \cite{navascues:2007,navascues:2008}, are not readily applicable to the PEF method. It is possible then that our results may have applications to other quantum information tasks for which a polytope approximation of the quantum set is desirable, and readers unfamiliar with the PEF method do not need to review it to understand the polytope approximations described in Sections \ref{s:background} and \ref{s:polytopes}. 

In the remainder of the paper, we review basic facts about the set of quantum-achievable probability distributions and related polytopes (Section \ref{s:background}), then derive our results about Tsirelson polytopes in Section \ref{s:polytopes}. In Section \ref{s:applications}, we review the PEF method and present a scenario in which using the new polytopes yields a demonstrable improvement over previous implementations, and then we finish with concluding remarks in Section \ref{s:conclusion}. An appendix contains some of the more technical proofs.

\section{Definitions and Background}\label{s:background}

Our setting is the (2,2,2) Bell scenario, in which there are two spatially separated parties (``Alice" and ``Bob") making measurements, two measurement settings for each party, and two possible measurement outcomes for each party. We can use random variables $O_A$ and $O_B$ to represent Alice's and Bob's outcomes, respectively, and random variables $S_A$ and $S_B$ to represent their respective settings. The outcome random variables $O_A$ and $O_B$ both take values in the set $\{0,+\}$ and the settings random variables $S_A$ and $S_B$ take values in the sets $\{a,a'\}$ and $\{b,b'\}$, respectively. A (2,2,2) Bell experiment is thus governed by a set of four conditional probability distributions corresponding to the four possible measurement configurations: $\{S_A=a\}\cap \{S_B=b\}$, $\{S_A=a\}\cap \{S_B=b'\}$, etc. We will use the shorthand where $P(\textnormal{0+}| a'b)$ is understood to mean $P(\{O_A=0\} \cap \{O_B=\textnormal+\}| \{S_A=a'\}\cap \{S_B=b\})$, and we follow the terminology of \cite{goh:2018}:
\begin{definition} 
A {\it behavior}, denoted $\vec P$, is a vector in $\mathbb R^{16}$ listing the set of 16 conditional probabilities $P(\textnormal{++}| ab)$, $P(\textnormal{+0}| ab)$, ... , $P(\textnormal{00}| a'b')$ corresponding to all possible measurement configurations and outcomes of the (2,2,2) experiment. A {\it Bell function} is a vector $\vec B$ in $\mathbb R^{16}$ inducing a function from behaviors to $\mathbb R$ via the dot product: $\vec B \cdot \vec P:\mathbb R^{16}\to \mathbb R$. 
\end{definition}
Bell functions will be used to introduce constraints of the form $\vec B \cdot \vec P\le R$, where $R$ is a real number, which will be satisfied only by some behaviors.

By the laws of probability, any valid behavior $\vec P$ must have only nonnegative entries and satisfy the following normalization equations:
\begin{eqnarray}
1 &=&P(\textnormal{++}| ab)+ P(\textnormal{+0}| ab) + P(\textnormal{0+}| ab)+ P(\textnormal{00}| ab) \notag\\
1 &=& P(\textnormal{++}| ab')+ P(\textnormal{+0}| ab') + P(\textnormal{0+}| ab')+ P(\textnormal{00}| ab')   \notag\\
1 &=& P(\textnormal{++}| a'b)+ P(\textnormal{+0}| a'b) + P(\textnormal{0+}| a'b)+ P(\textnormal{00}| a'b) \notag \\
1 &=& P(\textnormal{++}| a'b')+ P(\textnormal{+0}| a'b') + P(\textnormal{0+}| a'b')+ P(\textnormal{00}| a'b')  \label{e:prob}.
\end{eqnarray}
Furthermore, we will study only behaviors that additionally satisfy the {\it no-signaling constraints}:
\begin{eqnarray}
P(\textnormal{++}| ab)+ P(\textnormal{+0}| ab) &=& P(\textnormal{++}| ab')+ P(\textnormal{+0}| ab')\notag\\
P(\textnormal{++}| a'b)+ P(\textnormal{+0}| a'b) &=& P(\textnormal{++}| a'b')+ P(\textnormal{+0}| a'b')\notag\\
P(\textnormal{++}| ab)+ P(\textnormal{0+}| ab) &=& P(\textnormal{++}| a'b)+ P(\textnormal{0+}| a'b)\notag\\
P(\textnormal{++}| ab')+ P(\textnormal{0+}| ab') &=& P(\textnormal{++}| a'b')+ P(\textnormal{0+}| a'b')\label{e:nosig}
\end{eqnarray}
The no-signaling constraints express the condition that Alice's marginal outcome distribution should not depend on Bob's measurement choice and vice versa. For instance, the first equation above requires that when Alice's setting is $a$, her probability of getting a ``+" is the same whether Bob has setting $b$ or $b'$. The first equation also implies, when combined with the laws of probability expressed in \eqref{e:prob}, that Alice's probability of getting a ``0" does not depend on Bob's setting.

We will refer to the class of valid behaviors satisfying \eqref{e:nosig} as the no-signaling set $\mathcal {NS}$. $\mathcal {NS}$ forms a closed {\it convex polytope}: a bounded set formed by the intersection of finitely many closed half spaces. For $\mathcal{NS}$, the half spaces are obtained from the 8 linear equality constraints in \eqref{e:prob} and \eqref{e:nosig} combined with 16 linear inequality constraints ensuring that all entries are nonnegative, and the boundedness follows from the observation that $\mathcal{NS}\subseteq [0,1]^{16}$. It is well known \cite{barrett:2005} that $\mathcal {NS}$ can be expressed as the convex hull of 24 so-called {\it extreme points}, 8 of which are called ``Popescu-Rohrlich (PR) boxes" \cite{PRBOX} and 16 of which are called ``local deterministic" behaviors; see \cite{bierhorst:2016} Tables A1 and A2 for a list. We denote these extremal behaviors as $\{\vec{PR}_i\}_{i=1}^8$ and $\{\vec L_i\}_{i=1}^{16}$, respectively. The standard definition (such as in \cite{goh:2018}) of an {\it extreme point} of a set $\mathcal S$ is one that cannot be expressed as non-trivial convex combinations of other points in $\mathcal S$, and the Krein-Milman theorem implies that any convex compact set in $\mathbb R^n$ will be equal to the convex hull of its extreme points. This allows us to use the following working definition of a set of extreme points:

\newpage 

\begin{definition} \label{d:extreme}
Given a convex set $\mathcal S$, a subset $\mathcal E\subseteq \mathcal S$ is {\it the set of extreme points of $\mathcal S$} if 
\begin{enumerate}
\item $\mathcal S$ is contained in the convex hull of $\mathcal E$, denoted $\mathcal S \in \textnormal{Conv}(\mathcal E )$
\item No element of $\mathcal E$ can be expressed as a convex combination of other points in $\mathcal E$ 
\end{enumerate}
\end{definition}

There are two important subsets of $\mathcal {NS}$ to mention. First is the quantum set $\mathcal Q$, consisting of behaviors that can be induced by quantum measurements of a quantum system. Second is the local set $\mathcal L$, consisting of behaviors that admit a decomposition of the form $P(O_A,O_B|S_A,S_B) = \sum_\lambda P(O_A|S_A,\Lambda = \lambda)P(0_B|S_B,\Lambda=\lambda)P(\Lambda=\lambda)$ for a random variable $\Lambda$ that represents local hidden variables. $\mathcal L$ is equal to the convex hull of the 16 local deterministic distributions $\{\vec L_i\}_{i=1}^{16}$; recall these are some (but not all) of the extreme points of the set $\mathcal {NS}$. Quantum behaviors in $\mathcal{NS}\setminus \mathcal{L}$ can be shown to contain certifiable randomness.

Every Bell function has a maximum local value, a maximum quantum value, and a maximum no-signaling value which, given a Bell function $\vec B$, we define as
\begin{eqnarray}
LB &=& \sup_{\vec P \in \mathcal L}\vec B \cdot \vec P \quad \text{(Local Bound)}\notag\\
TB &=& \sup_{\vec P \in \mathcal Q}\vec B \cdot \vec P \quad \text{(Tsirelson Bound)}\notag\\
NSB &=& \sup_{\vec P \in \mathcal {NS}}\vec B \cdot \vec P \quad \text{(No-Signaling Bound)}\label{e:bounds}
\end{eqnarray}

For $\mathcal L$ and $\mathcal {NS}$, the above suprema are indeed maxima, as the maximum values are achieved. This follows because $\mathcal L$ and $\mathcal {NS}$ are each equal to the convex hull of a finite set of extreme points, and so for such a scenario, given a behavior $\vec P$ we can re-express $\vec B \cdot \vec P$ as 
\begin{equation*}
\vec B \cdot \vec P=\vec B \cdot \left ( \sum_{i=1}^n \lambda_i \vec E_i\right)=\sum_{i=1}^n \lambda_i \left(\vec B \cdot\vec E_i\right),
\end{equation*}
where the $\vec E_i$ are elements of the set of extreme points and the $\lambda_i $ are nonnegative numbers summing to one. Therefore no behavior can have a value of $\vec B \cdot \vec P$ that is greater than $\max_{\vec E_i\in\mathcal E}\vec B \cdot \vec E_i$. The fact that the maxima for $\mathcal L$ and $\mathcal {NS}$ are achieved at extreme points will be useful in the arguments below. Regarding the question of whether the supremum over the quantum set is a proper maximum, the situation is more complicated and discussed in \cite{goh:2018}, but this question is not material for the work below. Finally, it is well known that $\mathcal L \subseteq \mathcal Q \subseteq \mathcal{NS}$, and so the following inequality holds in general:
\begin{equation}\label{e:LBTBNSB}
LB \le TB \le NSB
\end{equation}
Thus Bell functions $\vec B$ for which $LB < TB$ holds strictly can be used to witness certifiable randomness in quantum behaviors $\vec P$ satisfying $\vec B \cdot \vec P > LB$.

One particularly important Bell function that we will discuss is the Clauser-Horne-Shimony-Holt (CHSH) Bell function $\vec B ^{\mathrm{CHSH}}$ \cite{CHSH}, whose coefficients are given in Table \ref{t:CHSH}. 
\begin{table}[h]\caption{The CHSH Bell function $\vec B ^{\mathrm{CHSH}}$ (left), a PR box behavior that achieves the no-signaling maximum $NSB=4$ of $\vec B ^{\mathrm{CHSH}}$ (center), and one of the eight local deterministic behaviors that achieves the local maximum $LB=2$ of $\vec B ^{\mathrm{CHSH}}$ (right). The entries of the table for $\vec B ^{\mathrm{CHSH}}$ give the coefficients that appear in the dot product $\vec B ^{\mathrm{CHSH}}\cdot\vec P$ defining the Bell function, so starting in the upper left, $1$ is the number to be multiplied by $P(\text{++}|ab)$, $-1$ is the number to be multiplied by $P(\text{+0}|ab)$, etc. The entries of the table for the PR box behavior are the probabilities themselves, so $P(\text{++}|ab)=1/2$, $P(\text{+0}|ab)=0$, etc. The entries of the table for the local deterministic behavior are also probabilities.}\label{t:CHSH}
\centering
\begin{tabular}{ c|rrrr }
 \multicolumn{1}{r}{}
  &  \multicolumn{1}{|r}{++}
 &  \multicolumn{1}{r}{+0}
 &  \multicolumn{1}{r}{0+} 
 &  \multicolumn{1}{r}{00}\\
  \cline{1-5}
$ab$ & $1$ & $-1$ & $-1$ & $1$\\
$ab'$ & $1$ & $-1$ & $-1$ & $1$\\
$a'b$ & $1$ & $-1$ & $-1$ & $1$\\
$a'b'$ & $-1$ & $1$ & $1$ & $-1$\\
\end{tabular}
\hspace{5mm}
\begin{tabular}{ c|cccc }
 \multicolumn{1}{r}{}
  &  \multicolumn{1}{|c}{++}
 &  \multicolumn{1}{c}{+0}
 &  \multicolumn{1}{c}{0+} 
 &  \multicolumn{1}{c}{00}\\
  \cline{1-5}
$ab$ & $1/2$ & $0$ & $0$ & $1/2$\\
$ab'$ & $1/2$ & $0$ & $0$ & $1/2$\\
$a'b$ & $1/2$ & $0$ & $0$ & $1/2$\\
$a'b'$ & $0$ & $1/2$ & $1/2$ & $0$\\
\end{tabular}
\hspace{5mm}
\begin{tabular}{ c|cccc }
 \multicolumn{1}{r}{}
  &  \multicolumn{1}{|c}{++}
 &  \multicolumn{1}{c}{+0}
 &  \multicolumn{1}{c}{0+} 
 &  \multicolumn{1}{c}{00}\\
  \cline{1-5}
$ab$ & $1$ & $0$ & $0$ & $0$\\
$ab'$ & $1$ & $0$ & $0$ & $0$\\
$a'b$ & $1$ & $0$ & $0$ & $0$\\
$a'b'$ & $1$ & $0$ & $0$ & $0$\\
\end{tabular}
\end{table}
The famous CHSH inequality, $\vec B ^{\mathrm{CHSH}} \cdot \vec P \le 2$ for $\vec P \in \mathcal L$, is a statement that $LB$ is 2 for $\vec B ^{\mathrm{CHSH}}$. The Tsirelson bound $TB$ for $\vec B ^{\mathrm{CHSH}}$ is $2\sqrt 2$ \cite{tsirelson:1993}. The no-signaling bound $NSB$ of $4$ is achieved by the PR box behavior in Table \ref{t:CHSH}  \cite{PRBOX}. The local bound $LB$ of 2 is achieved by eight local deterministic behaviors; each of these eight behaviors has a entry of ``1" in exactly one place where the PR box has a ``0," and this location is unique to each of the eight saturating local behaviors.

There are symmetries of the convex sets $\mathcal{NS}$, $\mathcal Q$, and $\mathcal L$, for which the associated transformations applied to the CHSH inequality generate new inequalities. There are eight inequivalent versions of the CHSH inequality obtained this way. Each version of the CHSH inequality corresponds to a unique PR box behavior obtaining the $NSB$ of $4$ with a corresponding set of eight local deterministic behaviors obtaining the $LB$ of 2.

\section{Tsirelson Polytopes}\label{s:polytopes}

Given a Bell function $\vec B$ and a real number $TB^*$ satisfying $TB\le TB^* \le NSB$, we define the corresponding {\it Tsirelson Polytope} $\mathcal{Q}_T$ to be $\{\vec P \in \mathcal {NS} | \vec B \cdot \vec P \le TB^*\}$. We allow for $TB^*$ to exceed the quantum supremum $TB$, because we might want to consider Bell functions for which the exact quantum limit is not known but a numerical upper bound can be found \cite{navascues:2007,navascues:2008}. As $\mathcal{Q}_T$ is the intersection of $\mathcal {NS}$ with a half space defined by a linear inequality, $\mathcal Q_T$ will form a polytope.

Not all Bell functions will lead to scenarios worth studying. For our purposes, any interesting Bell function should not have $NSB=LB$, which by \eqref{e:LBTBNSB} would lead to the degeneracy $LB = TB = NSB$, so we only consider Bell functions for which $LB <NSB$ holds strictly. For a given Bell function $\vec B$, one can effectively determine whether the strict inequality holds by checking the value of $\vec B \cdot \vec E$ for all extreme points of $\mathcal L$ and $\mathcal {NS}$. Here is a useful lemma about such Bell functions:
\begin{lemma}\label{f:exactlyone}
For any Bell function $\vec B$ for which $LB < NSB$ holds, there is exactly one PR box for which $\vec B \cdot \vec{PR}>LB$, and $\vec B \cdot \vec{PR} = NSB$ for this PR box.
\end{lemma}
\emph{Proof.}
$NSB$ will be achieved at an extreme point of $\mathcal {NS}$, and the assumption $LB < NSB$ implies that this extreme point must be one of the PR boxes. If a second PR box satisfied $\vec B \cdot \vec{PR}>LB$, then an equal mixture of these two PR boxes -- which is in the local set $\mathcal L$ (see Theorem 2.1 in \cite{bierhorst:2016}) -- would exceed the local bound, a contradiction.
$\hfill\Box$

\medskip

A priori, any Bell function satisfying $LB <NSB$ will fall into one of two categories.: $LB\le TB < NSB$ and $LB < TB =  NSB$. It turns out the latter of these is impossible: Appendix D of \cite{goh:2018} explains that if $TB = NSB$, then $LB = TB = NSB$. Thus we need only consider Bell functions for which $LB\le TB < NSB$; this forms our general scenario of interest. For the rest of this paper, we assume this condition, as well as $TB^*< NSB$ (as $TB^*=NSB$ yields $\mathcal Q_T = \mathcal {NS}$).

Lemma \ref{f:exactlyone} tells us that when we intersect $\mathcal{NS}$ with $\{\vec P | \vec B \cdot \vec P \le TB^*\}$ for some $TB^*\in [TB,NSB)$, the resulting set contains all of the extreme points of $\mathcal{NS}$ save one -- the sole PR box for which $\vec B \cdot \vec{PR} =NSB>LB$. Thus many of the extreme points of $\mathcal {NS}$ are extreme points of $\mathcal Q_T$. The following theorem, proved in the appendix, classifies the rest of the extreme points of $\mathcal Q_T$.

\begin{theorem}\label{t:classify}
Let $\vec B$ be a Bell function for which $LB\le TB < NSB$ holds, and let $\vec{PR}_1$ denote the PR box for which $\vec B\cdot \vec{PR}_1=NSB$. Let $TB^*\in [TB,NSB)$ and define $\mathcal Q_T= \{\vec P \in \mathcal{NS}|\vec B \cdot \vec P \le TB^*\}$. Then the set of extreme points of $\mathcal Q_T$ is equal to the set $\mathcal E$ defined as follows: all of the local deterministic distributions, all of the PR boxes except $\vec {PR}_1$, and all behaviors of the form
\begin{equation}\label{e:newextreme}
\vec E_i = \lambda_i\vec{PR}_1 + (1-\lambda_i)\vec{L}_i 
\end{equation}
where $\vec L_i$ is one of the eight local distributions saturating the version of the CHSH inequality maximally violated by $\vec{PR}_1$, and $\lambda_i=(TB^*-B_i)/(NSB-B_i)$ with $B_i = \vec B \cdot \vec L_i$.
\end{theorem}
We note that if $LB=TB^*$ holds, a behavior defined by \eqref{e:newextreme} can coincide with a local deterministic distribution, in which case the statement of the theorem refers to this behavior twice in defining the set $\mathcal E$.

Applying the above theorem to the special case of the original Tsirelson bound of $2\sqrt 2$ for $\vec B ^{\mathrm{CHSH}}$, one obtains the same value of $\lambda_i = \sqrt 2 - 1$ for all eight versions of the $\vec E_i$ behavior in \eqref{e:newextreme}. Furthermore, if one simultaneously introduces all eight versions of the CHSH inequality with Tsirelson bounds $2\sqrt 2$, it follows from the above arguments that each version causes the corresponding PR box behavior achieving $NSB=4$ to split into eight extreme points of the form \eqref{e:newextreme}, resulting in a polytope with 80 extreme points, as described in \cite{PEF}.

The scenario of the 80-vertex polytope is straightforward to describe because the different Tsirelson bounds do not ``interact" in the sense that the respective no-signaling maxima for their corresponding Bell functions are achieved by different PR box behaviors; this is depicted schematically in Figure \ref{f:noninteracting}. However, the situation is more complicated if Tsirlson bounds are introduced for two distinct Bell functions maximized by the same PR box, as represented in Figure \ref{f:schematic}. This turns out to be the scenario that yields improvement for randomness certification.
\begin{figure}[h]\caption{{\bf Schematic diagrams of the space of behaviors for a (2,2,2) Bell scenario.} The true space is 8-dimensional; these figures represents a 2-dimensional slice of that space. Behaviors in the innermost $\mathcal L$ region cannot be used for device-independent applications. Behaviors in the curved $\mathcal Q$ region are quantum-achievable. The $\mathcal Q$ region is contained by the no-signaling polytope $\mathcal{NS}$.}
\begin{subfigure}{0.48\textwidth}
\begin{tikzpicture}
\def\sca{2.6}
\draw (-\sca*0.7071-\sca,0) node [label={[label distance=\sca*1.5]3:$\small\mathcal L$}]{} --(-\sca*0.7071-\sca,\sca*0.7071)--(\sca*0.7071-\sca,\sca*0.7071)--(\sca*0.7071-\sca,-\sca*0.7071)--(-\sca*0.7071-\sca,-\sca*0.7071)--(-\sca*0.7071-\sca,0);
\draw (-\sca*0.7071*2-\sca,0)  node [label={[label distance=\sca*1.2]3:$\small\mathcal{NS}$}] {} --(-\sca,2*\sca*0.7071) --(\sca*0.7071*2-\sca,0) -- (-\sca,-2*\sca*0.7071) -- (-\sca*0.7071*2-\sca,0);
\draw [dashed] (-.75*\sca-\sca,\sca)--(.75*\sca-\sca, \sca) ;
\draw [dashed] (-.75*\sca-\sca,-\sca)--(.75*\sca-\sca, -\sca) ;
\draw [dashed] (0,.75*\sca)--(0, -.75*\sca) ;
\draw (0,0) arc (0:180:\sca)node [label={[label distance=\sca*.5]3:\small$\mathcal Q$}] {};
\draw (0,0) arc (0:-180:\sca);
\end{tikzpicture}\caption{{\it Non-interacting Tsirelson bounds.} The dashed lines represent three different versions of the original Tsirelson bound. When a single such bound is incorporated, the resulting polytope replaces a PR box behavior, located at a vertex of the larger square, with extreme points of the form \eqref{e:newextreme}. These new extreme points can be thought of schematically as occurring where a dashed line intersects the outer square, though the actual location will not be contained in the 2-dimensional slice depicted.
}\label{f:noninteracting}
\end{subfigure}
\hspace*{\fill} 
\begin{subfigure}{0.48\textwidth}
\begin{tikzpicture}
\def\ra{2.6}
\def\sca{2.6}
\draw (-\sca*0.7071-\sca,0) node [label={[label distance=\sca*1.5]3:$\small\mathcal L$}]{} --(-\sca*0.7071-\sca,\sca*0.7071)--(\sca*0.7071-\sca,\sca*0.7071)--(\sca*0.7071-\sca,-\sca*0.7071)--(-\sca*0.7071-\sca,-\sca*0.7071)--(-\sca*0.7071-\sca,0);
\draw (-\sca*0.7071*2-\sca,0)  node [label={[label distance=\sca*1.2]3:$\small\mathcal{NS}$}] {} --(-\sca,2*\sca*0.7071) --(\sca*0.7071*2-\sca,0) -- (-\sca,-2*\sca*0.7071) -- (-\sca*0.7071*2-\sca,0);
\draw [dotted,thick] (-\ra-0.7071*\ra*1.3,\ra*1.0824-.2829*\ra*1.3) node [left] {\footnotesize\eqref{e:tiltineq}}--(-\ra+0.7071*\ra*.2,\ra*1.0824+.2829*\ra*.2);
\draw [dashed] (-.95*\ra-\ra,\ra)--(.95*\ra-\ra, \ra) node [above] {\footnotesize Original Tsirelson};
\draw (0,0) arc (0:180:\sca)node [label={[label distance=\sca*.5]3:\small$\mathcal Q$}] {};
\draw (0,0) arc (0:-180:\sca);
\end{tikzpicture}\caption{{\it Tsirelson bounds that interact.} The uppermost portion of the $\mathcal Q$ region can be approximated to increasing degrees of accuracy by incorporating the original Tsirelson bound (dashed line) in conjunction with a bound of the ``tilted'' form \eqref{e:tiltineq} with $\alpha>1$ (dotted line). The extreme points given by expression \eqref{e:doubleextreme} can be thought of schematically as occurring where the dashed and dotted lines intersect. {\color{white} aaaaaaaaaa bbbbbbbbbb ccccccccccc ddddd eeeee fffff ggggg hhhhh iiiii }}\label{f:schematic} 
\end{subfigure}
\end{figure}
With a little effort, the proof method for Theorem \ref{t:classify} can be adapted to find the extreme points of the more complicated polytopes with two Tsirelson bounds interacting non-trivially.

As an example, let us consider the ``tilted" CHSH Bell function $\vec B^\alpha$ whose coefficients are given in Table \ref{t:tiltedCHSH}. 
\begin{table}[t]\caption{The ``tilted" CHSH Bell function $\vec B^\alpha$ for values of $\alpha>1$. This reduces to $\vec B ^{\mathrm{CHSH}}$ when $\alpha=1$.}\label{t:tiltedCHSH}
\centering
\begin{tabular}{ r|rrrr }
 \multicolumn{1}{r}{}
  &  \multicolumn{1}{|r}{++}
 &  \multicolumn{1}{r}{+0}
 &  \multicolumn{1}{r}{0+} 
 &  \multicolumn{1}{r}{00}\\
  \cline{1-5}
$ab$ & $\alpha$ & $-\alpha$ & $-\alpha$ & $\alpha$\\
$ab'$ & $\alpha$ & $-\alpha$ & $-\alpha$ & $\alpha$\\
$a'b$ & $1$ & $-1$ & $-1$ & $1$\\
$a'b'$ & $-1$ & $1$ & $1$ & $-1$\\
\end{tabular}
\end{table}
The Tsirelson bound for this Bell function is derived in Ref.~\cite{acin:2012}, valid for all values of $\alpha>1$:
\begin{equation}\label{e:tiltineq}
\vec B^\alpha\cdot \vec P \le 2\sqrt{1+\alpha^2} \quad \text{ for all } \vec P \in \mathcal Q
\end{equation}
There is a quantum behavior saturating the bound. For a fixed $\alpha >1$, the extreme points of the polytope induced by \eqref{e:tiltineq} alone are given by Theorem \ref{t:classify}. Now let us consider the polytope of behaviors that obey Tsirelson's original inequality $\vec B ^{\mathrm{CHSH}}\cdot \vec P \le 2\sqrt 2$ as well as \eqref{e:tiltineq} for a fixed value of $\alpha>1$, as depicted schematically in Figure \ref{f:schematic}.

To describe the extreme points of this polytope, first we remark that of the eight local deterministic behaviors satisfying $\vec B ^{\mathrm{CHSH}}\cdot \vec L=2$, four of them will have a $\vec B^\alpha\cdot \vec L $ value of $2$, and four of them will have a $\vec B^\alpha\cdot \vec L $ value of $2\alpha$. This is confirmed by inspection of Table \ref{t:tiltedCHSH}. We can label the first four of these local deterministic behaviors $\{\vec L^{\mathrm{top}}_i\}_{i=1}^4$, and the second four $\{\vec L^{\mathrm{bot}}_i\}_{i=1}^4$. 

We assert that the extreme points of the polytope are those for $\mathcal {NS}$, minus the single PR box that violates both Tsirelson bounds, plus 24 new extreme points. Four of these extreme points are obtained from the expression \eqref{e:newextreme} using $\vec L^{\mathrm{top}}_i$ vectors with $NSB$, $B_i$, and $TB^*$ generated by $\vec B ^{\mathrm{CHSH}}$, and four are obtained from the expression \eqref{e:newextreme} using the $\vec L^{\mathrm{bot}}_i$ vectors with $NSB$, $B_i$, and $TB^*$ generated by $\vec B^\alpha$. Note these behaviors each saturate one of the two Tsirelson inequalities while strictly obeying the other one. The remaining 16 extreme points, which saturate both Tsirelson inequalities, are given as
\begin{equation}\label{e:doubleextreme}
\lambda_{PR} \vec{PR}_1 + \lambda_{\mathrm{top}}\vec L^{\mathrm{top}}_i+ \lambda_{\mathrm{bot}}\vec L^{\mathrm{bot}}_j
\end{equation}
for all choices of $i,j \in \{1,2,3,4\}$, where the $\lambda$ coefficients are found by solving the simultaneous set of equations
\begin{eqnarray*}
(2+2\alpha)\lambda_{PR}  + 2\lambda_{\mathrm{top}}+ 2\alpha\lambda_{\mathrm{bot}}&=&2\sqrt{1+\alpha^2}\\
4\lambda_{PR}  + 2\lambda_{\mathrm{top}}+ 2\lambda_{\mathrm{bot}}&=&2\sqrt{2}\\
\lambda_{PR}  + \lambda_{\mathrm{top}}+ \lambda_{\mathrm{bot}}&=&1.
\end{eqnarray*}

\noindent The following values solve this set of equations:

\begin{equation}\label{e:coeffs}
\lambda_{PR}=\sqrt{2}-1\quad\quad\quad
\lambda_{\mathrm{top}}=1-\frac{\sqrt{1+\alpha^2}-\sqrt{2}}{\alpha-1}\quad\quad\quad
\lambda_{\mathrm{bot}}=1-\frac{\alpha\sqrt{2}-\sqrt{1+\alpha^2}}{\alpha-1}
\end{equation}
Furthermore, the condition $\alpha>1$ ensures the above coefficients are nonnegative,\footnote{The subtracted fractions in the expressions for $\lambda_{\mathrm{top}}$ and $\lambda_{\mathrm{bot}}$ are less than one:\newline
$\alpha>1 \Rightarrow 1+\alpha^2 +(2\sqrt{2}-2)(\alpha-1)>1+\alpha^2
\Rightarrow (\alpha -1 +\sqrt{2})^2>1+\alpha^2
\Rightarrow \alpha -1 >\sqrt{1+\alpha^2}-\sqrt{2}$, and \newline
$\alpha>1 \Rightarrow 1+\alpha^2-(2\sqrt{2}-2)\alpha(\alpha-1)<1+\alpha^2
\Rightarrow [(\sqrt{2}-1)\alpha+1]^2<1+\alpha^2
\Rightarrow \alpha\sqrt{2}-\sqrt{1+\alpha^2}<\alpha-1$.
}
 so the expression \eqref{e:doubleextreme} with the coefficients in \eqref{e:coeffs} yields a valid convex combination. An outline of the proof that these are the extreme points is given in the appendix.

\section{Applications to Device Independent Random Number Generation}\label{s:applications}

The constructions of Section \ref{s:polytopes} can be applied to the task of device-independent random number generation via the method of probability estimation factors \cite{zhang:2018,PEF}, which we now briefly review. In the context of device-independent randomness, a {\it probability estimation factor} (PEF)  is a function, satisfying certain conditions, that maps the result of a Bell experiment to the nonnegative real numbers. It is essentially a score function that assigns higher values to more nonlocal results, thereby quantifying certifiable randomness. In our (2,2,2) Bell scenario, the result of an experimental trial consists of the two settings choices and two outcomes for Alice and Bob, so any PEF is a function $F: (O_A,O_B,S_A,S_B)\to \mathbb R^+\cup \{0\}$ with sixteen possible inputs. The precise definition of a PEF in this scenario is as follows: for a collection of behaviors $\mathcal P$ and a parameter $\beta>0$, a {\it PEF with power $\beta$} is a nonnegative function satisfying the following inequality for all behaviors $\vec P\in \mathcal P$:
\begin{equation}\label{e:PEFdef}
E_{\vec P}\left [F(O_A,O_B,S_A,S_B)P(O_A,O_B|S_A,S_B)^\beta\right]\le 1,
\end{equation}
where $E_{\vec P} [\cdot ]$ denotes expected value with respect to the joint distribution $P(O_A,O_B,S_A,S_B)$ of the settings and outcomes given by the behavior $\vec P$ with a fixed settings distribution. The probability $P(O_A,O_B|S_A,S_B)$ in \eqref{e:PEFdef} is also according to the behavior $\vec P$. 

Equation \eqref{e:PEFdef} can be used to prove that a certain function of the cumulative value of PEF scores over a sequence of $n$ repeated trials of a Bell experiment is a bound on the probability of the string of observed outcomes, and thus a witness to the presence of randomness. Specifically, in \cite{zhang:2018} it is shown that if $F$ is a PEF with power $\beta$ for a collection of behaviors $\mathcal P$, then for any $\epsilon>0$, the following holds for any probability distribution $P$ governing all $n$ trials in which the trial-by-trial, settings-conditional outcome distributions are (possibly different) behaviors in $\mathcal P$:
\begin{equation}\label{e:probbound}
P\left[P(\text{outcome string}|\text{setting string})\ge \left(\epsilon\prod_{i=1}^nF_i\right)^{-1/\beta}\right]\le \epsilon.
\end{equation}
The above expression, roughly speaking, says that when the product of the trial-by-trial PEF values $F_i$ is large  -- and so the quantity $\left(\epsilon\prod_{i=1}^nF_i\right)^{-1/\beta}$ is small -- it is unlikely (outer probability less than epsilon) that the observed outcome string occurs with more than a small probability (inner inequality). Often, we assume that the collection of behaviors $\mathcal P$ is the set of quantum-achievable behaviors, and so \eqref{e:probbound} will hold under the assumption that quantum mechanics is correct. Importantly, it is not necessary to assume that that the $n$ trials are independent and identically distributed (i.i.d.); the behavior can vary within $\mathcal P$ from trial to trial and \eqref{e:probbound} will still hold. Equation \eqref{e:probbound} will only be useful in the presence of nonlocal behaviors: barring an unlikely statistical fluctuation, classical behaviors admitting a local realistic description will yield a trivial probability bound $\left(\epsilon\prod_{i=1}^nF_i\right)^{-1/\beta}$ in the sense that the bound will be greater than or equal to 1.

Polytopes containing the quantum set of behaviors $\mathcal Q$ can be used to effectively construct valid PEFs satisfying the defining constraint \eqref{e:PEFdef}. Specifically, Section V of the supplemental material of \cite{zhang:2018} shows that anything satisfying the PEF defining condition \eqref{e:PEFdef} for the extreme points of a polytope containing $\mathcal P$ will satisfy \eqref{e:PEFdef} for all the behaviors in $\mathcal P$. Hence it becomes possible to check that a PEF is valid by checking only a finite number of linear inequality constraints, one for each extreme point.

The probability bound of $\left(\epsilon\prod_{i=1}^nF_i\right)^{-1/\beta}$ in \eqref{e:probbound} is a rough measure of how much randomness is available in the data of a Bell experiment. To illustrate, suppose an experiment is run and $\left(\epsilon\prod_{i=1}^nF_i\right)^{-1/\beta}$ comes out to equal, say, $(1/2)^{26}$. Then \eqref{e:probbound} implies that, modulo an error probability of $\epsilon$, the likelihood of the observed outcome string was no more than the likelihood of observing a particular 26-bit string drawn from a uniform stream of random bits. However, the actual outcome string of the raw data of the Bell experiment may be thousands of bits long; to process the randomness in this long string into a shorter, near-uniform output string for information-theoretic applications, further work can be done \cite{zhang:2018,bierhorst:2018,zhang:2020,PEF} to account for the non-unity probability that the PEF product exceeds a certain threshold and to properly apply classical postprocessing machinery such as an extractor function \cite{trevisan:2001}. These steps consume some of the randomness and would result in a final output string of fewer than 26 bits. However, we will not implement these final steps and will just use the probability bound of $\left(\epsilon\prod_{i=1}^nF_i\right)^{-1/\beta}$ in \eqref{e:probbound} as a measure of available certifiable randomness for comparing performance of various PEFs. This is reasonable because in the limit of a large number of i.i.d. trials sampled from a nonlocal behavior, the negative base-2 logarithm of the probability bound $\left(\epsilon\prod_{i=1}^nF_i\right)^{-1/\beta}$ converges asymptotically to the number of extractable bits \cite{zhang:2018}. We use the negative base-2 logarithm scale when reporting results.

Given a quantum-achievable behavior, we can compare the performance of PEFs for it by fixing an error bound $\epsilon$, fixing a number $n$ of i.i.d.~trials sampled from the quantum behavior with equiprobable settings, and computing $\left(\epsilon\prod_{i=1}^nF_i\right)^{-1/\beta}$ in \eqref{e:probbound}. Since $\prod_{i=1}^nF_i$ is a random variable, the probability bound will depend on the particular instance of the experiment. However, we can anticipate a likely value for $\prod_{i=1}^nF_i$ using the fact that for sufficiently large $n$, $\sum_{i=1}^n\ln \left ( F_i \right)$ will be either greater than or less than $n E(\ln F)$ with roughly equal probability, where the expectation $E$ is computed according to the chosen quantum behavior with equiprobable settings. (This follows from the Central Limit Theorem; see \cite{zhang:2011}, or \cite{bierhorst:2018} Supplementary Information Section 3 for details.) Hence we use 
\begin{equation}\label{e:likelyvalue}
\left\{\epsilon\exp\left [n E(\ln F)\right]\right\}^{-1/\beta},
\end{equation}
where $\exp$ denotes the exponential function, as the median anticipated value for $\left(\epsilon\prod_{i=1}^nF_i\right)^{-1/\beta}$. Since it is desirable for the probability bound $\left(\epsilon\prod_{i=1}^nF_i\right)^{-1/\beta}$ in \eqref{e:probbound} to be small (lower probabilities correspond to more randomness), finding a PEF that minimizes \eqref{e:likelyvalue} will optimize performance.

\begin{figure}[t!]\centering
\includegraphics[scale=0.8]{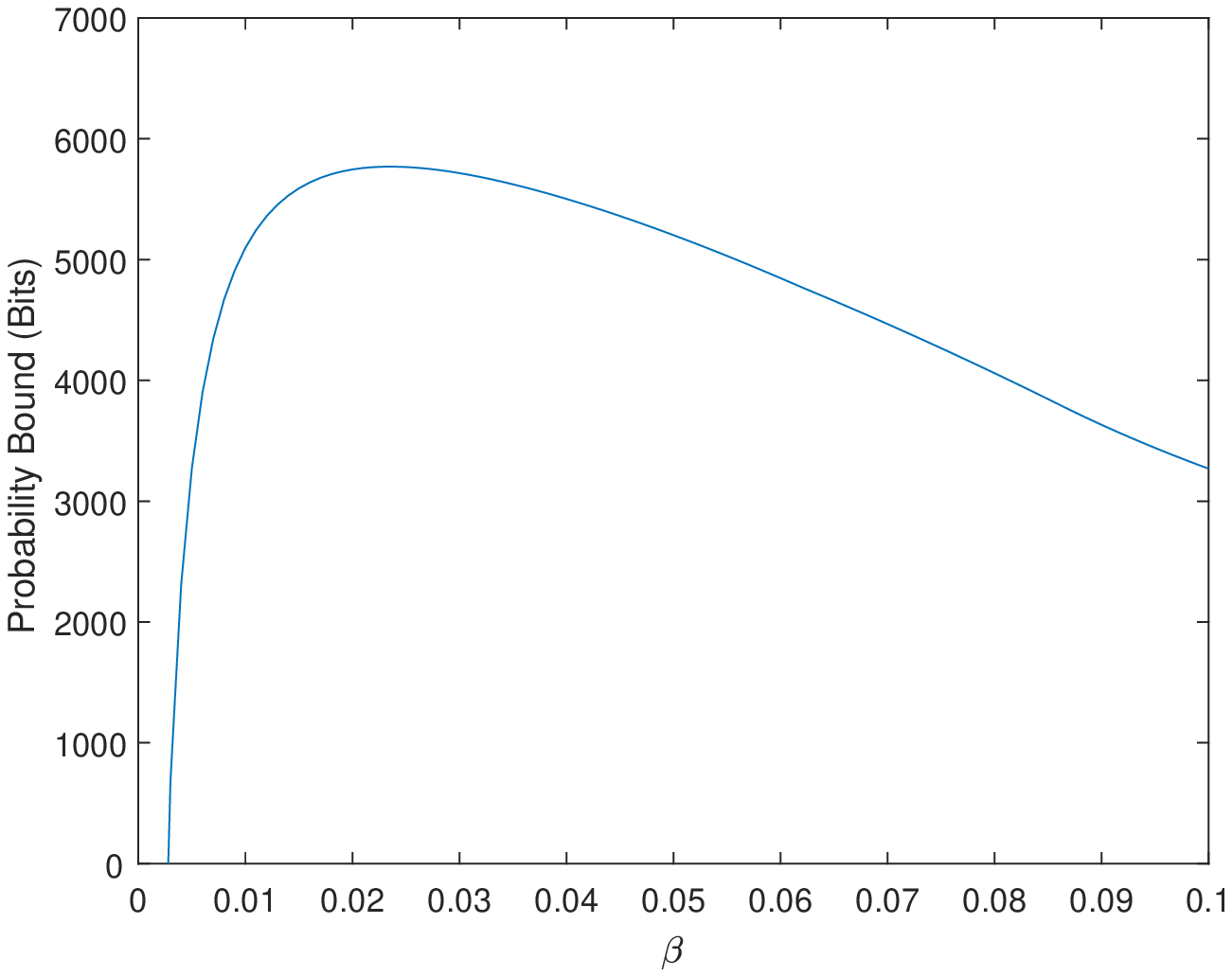}
\caption{Optimizing \eqref{e:likelyvalue} by performing the maximization problem of \eqref{e:maximization} for various values of $\beta$, with $\epsilon=10^{-6}$ and $n = 10,000$. The distribution of the trial results is given by the unique quantum behavior $\vec P$, given in \cite{acin:2012}, that saturates \eqref{e:tiltineq} with $\alpha=2$. The polytope used to generate the extreme points $\mathcal E$ in \eqref{e:maximization} is induced by (only) the original Tsirelson inequality $\vec B^{\mathrm{CHSH}}\cdot \vec P\le 2\sqrt 2$. The Y-axis plots $-\log_2$ of the quantity in \eqref{e:likelyvalue}, to report the median anticipated value of the probability bound $\left(\epsilon\prod_{i=1}^nF_i\right)^{-1/\beta}$ in bits. A similar phenomenon is observed in Figure 1 of \cite{PEF}.}\label{f:betacurve}
\end{figure}

Given a set of extreme points $\mathcal E$ of a polytope containing the quantum set $\mathcal Q$, we can perform a maximization procedure to find a PEF, valid for all quantum behaviors, that optimizes the quantity \eqref{e:likelyvalue} for a fixed number of trials $n$ and a fixed error parameter $\epsilon$. The procedure is as follows: first fix a choice of $\beta$, and note that with $n$, $\epsilon$, and $\beta$ fixed, \eqref{e:likelyvalue} is a monotone decreasing function of $E(\ln F)$. Now, in order to minimize \eqref{e:likelyvalue}, we need to solve the following convex optimization problem:
\begin{eqnarray}\label{e:maximization}
\underset{F}{\text{Maximize}} &&  E(\ln F)\\
\text{Subject to} && E_{\vec E}\left [F(O_A,O_B,S_A,S_B)P_{\vec E}(O_A,O_B|S_A,S_B)^\beta\right]\le 1, \quad \forall \vec E \in \mathcal E\notag\\
&& F(O_A,O_B,S_A,S_B)\ge 0, \quad \forall O_A,O_B,S_A,S_B.\notag
\end{eqnarray}
This can be solved effectively with standard computer algorithms. Once this is done, plug the resulting optimized value of $E(\ln F)$ into \eqref{e:likelyvalue}. Then repeat the procedure for various values of $\beta$ looking for the lowest possible value of \eqref{e:likelyvalue}. By numerics, we found that the optimal value of $\beta$ for each example studied in this work lies in the interval [0.001, 0.100]. To obtain the lowest possible value of \eqref{e:likelyvalue} and the corresponding optimal value of $\beta$, we tested 100 equally spaced values of $\beta$ between 0.001 and 0.100, which yields a curve with a clear optimum as illustrated in Figure \ref{f:betacurve}.

\medskip

\noindent {\it Results.} To demonstrate the method, we considered the amount of randomness certifiable from a specific quantum-achievable behavior. Our choice was the unique quantum behavior, given in \cite{acin:2012}, that maximizes the Bell function in Table \ref{t:tiltedCHSH} when $\alpha=2$. Since this Bell function uniquely determines the behavior, a reasonable conjecture might be that the polytope obtained from the corresponding Tsirelson bound \eqref{e:tiltineq} with $\alpha=2$ would result in a larger amount of randomness than could be obtained using the polytope induced by the original Tsirelson bound $\vec B^{\mathrm{CHSH}}\cdot \vec P \le 2\sqrt 2$ (equivalent to \eqref{e:tiltineq} with $\alpha=1$). We found that the opposite was true. However, including both bounds to create a smaller polytope does result in a meaningful improvement compared to using either bound individually. Note that \cite{zhang:2018,zhang:2020,PEF,shalm:2019,li:2019} all use the polytope with only $\vec B^{\mathrm{CHSH}} \le 2\sqrt 2$, so our observation that the polytope using both bounds results in an improvement of almost 18\% to the number of certified random bits is relevant for future implementations. Our results are summarized in Table \ref{t:results1}.

\begin{table}[t]\caption{Comparison of the amount of randomness obtained from three different polytopes with $\epsilon=10^{-6}$ and $n = 10,000$. The ``Probability Bound (Bits)" quantity is obtained by calculating {\footnotesize $\underset{\beta\in (0.001,0.1)}{\text{max}} \, \underset{F}{\text{max}} \, -\log_2\left(\left\{\epsilon\exp\left [n E(\ln F)\right]\right\}^{-1/\beta}\right)$}, where the expectation is computed for the quantum distribution $\vec P$ that maximizes $\vec B^\alpha \cdot \vec P$ for $\alpha=2$. The inner maximum is calculated according to \eqref{e:maximization}, and then recalcuated for 100 equally spaced values of $\beta\in(0.001,0.1)$ to obtain the outer maximum. }\label{t:results1}
\centering
\begin{tabular}{ cccc }
& \eqref{e:tiltineq}, $\alpha=2$ & Original Tsirelson  &  Both\\ \toprule
Probability Bound (Bits) & 5,187.65 & 5,769.10 & 6,805.23\\
\end{tabular}
\end{table}

\begin{figure}[t]\centering
\includegraphics[scale=0.8]{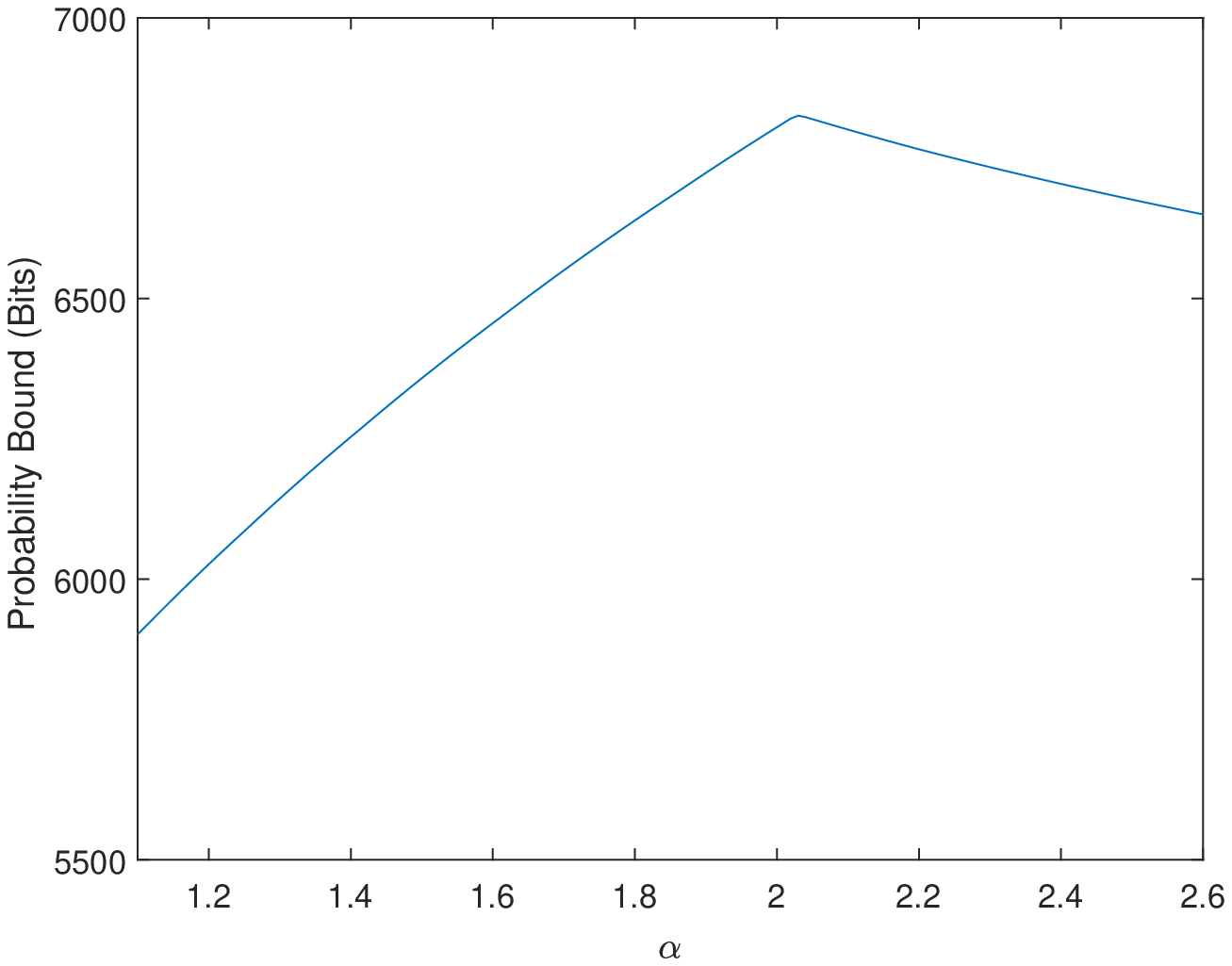}
\caption{For various choices of $\alpha$, we compute the polytope induced by both \eqref{e:tiltineq} and $\vec B^{\mathrm{CHSH}} \cdot \vec P \le 2\sqrt 2$. We then perform the optimization problem \eqref{e:maximization} using this polytope and the distribution used in Table \ref{t:results1}. The optimization problem is performed multiple times with multiple values of $\beta\in(0,0.1)$. For each choice of $\alpha$ we report the largest value of $-\log_2$ of \eqref{e:likelyvalue} found for all $\beta$.}\label{f:alphacurve}
\end{figure}

We are of course not limited to the three polytopes analyzed in Table \ref{t:results1}. We have derived a parametrized family of two-inequality polytopes for different values of $\alpha$. Figure \ref{f:alphacurve} displays interesting behavior for the amount of certifiable randomness at different values of $\alpha$, and supports the notion that the $\alpha=2$ figure of $6,805.23$ reported in Table \ref{t:results1} is near-optimal for this family of polytopes. Numerical evidence indicates that the optimal value of $\alpha$ is closer to $2.03$. The abrupt cusp in the curve at this maximum is a striking feature, especially as the extreme points $\vec E_i$ given by \eqref{e:newextreme} and \eqref{e:doubleextreme} vary continuously as a function of $\alpha$. We speculate it may indicate that the restrictions imposed by the constraints in the PEF optimization \eqref{e:maximization} are dominated by one particular $\vec E_i$ and the cusp occurs at a location where there is a switch in which $\vec E_i$ dominates. Exploring the causes and prevalence of this type of feature could be an aspect of future work.

There are many different scenarios that can also be examined. The Tsirelson bound for expression (17) of Ref.~\cite{goh:2018}, computed using the analytic technique of Wolfe and Yelin \cite{wolfe:2012}, is saturated by a quantum behavior as well as a local deterministic behavior -- and consequently is saturated by any convex mixture of these two behaviors. We found that the amount of randomness that can be certified from such a convex mixture is greater using the standard CHSH inequality polytope, compared to what can be certified using the polytope induced by (17) in Ref.~\cite{goh:2018}, paralleling our finding above for the single-inequality polytopes induced by $\vec B^\alpha$ and $\vec B^{\mathrm{CHSH}}$. Furthermore, initial explorations employing polytopes with two non-trivially interacting Tsirelson bounds to certify randomness in existing experimentally generated data sets, such as the data in Ref.~\cite{zhang:2020}, produced only marginal improvements over the results reported in the reference (which uses only the original Tsirelson inequality). It may be that the meaningful improvements reported in Table \ref{t:results1} are more characteristic of behaviors near the quantum boundary and/or behaviors inducing a large absolute violation of the original CHSH inequality. This is worth considering if loophole-free Bell experiments employing atoms \cite{hensen:2015,rosenfeld:2017}, which achieve higher violations of the CHSH inequality than photonic loophole-free Bell experiments \cite{shalm:2015,giustina:2015}, begin to reach the higher data rates characteristic of photonic experiments. To date, only photonic systems have reached data rates sufficient to enable practical demonstrations of device-independent random number generation \cite{bierhorst:2018,liu:2018,zhang:2020,shalm:2019,li:2019}.

\section{Conclusion}\label{s:conclusion}

We have derived formulas for identifying the extreme points of polytopes induced by multiple Tsirelson bounds, relating the structure of these polytopes to parameters in the defining inequalities. We have demonstrated that these results can be used to improve the performance of the probability estimation factor method for certifying device-independent randomness. Our techniques outline a general approach for classifying the extreme points of such polytopes, and in future work it may be useful to incorporate three or more non-trivially interacting Tsirelson bounds to obtain better approximations of the quantum set. Unfortunately, the complexity of a polytope grows with each additional constraint and a law of diminishing returns will apply as each iteration removes a smaller volume of behaviors than the one before it. Ideally the limiting behavior of performance enhancements will become apparent before the procedure becomes intractable. Even small improvements can be worth pursuing: in current device-independent random number generation experiments \cite{bierhorst:2018,liu:2018,zhang:2020,shalm:2019,li:2019}, the speed of randomness generation tops out at kilobits per second, whereas even a partial relaxation of assumptions such as in semi-device independent protocols with an uncharacterized source and characterized measurements \cite{cao:2016,marangon:2017,xu:2019} can yield generation rates in the range of gigabits per second. This highlights how improvements to the amount of randomness on the scale reported in this paper could be quite valuable in applications where the full security of a completely device-independent implementation, with its attendant reduction in randomness generation rate, is demanded. The results presented here may also find applications beyond probability estimation, given the growing scope of quantum information theory and in particular device-independent protocols. 

\medskip

\section*{Acknowledgements} The authors thank Emanuel Knill for helpful input on this project. This work was partially supported by National Science Foundation award number 1839223 and the Louisiana Board of Regents contract number LEQSF(2019-22)-RD-A-27.

\appendix
\section{Proofs of Results}

In this appendix, we prove Theorem \ref{t:classify}, and then outline the proof of the classification of the extreme points of the polytope induced by $\vec B ^{\mathrm{CHSH}}$ and $\vec B ^\alpha$.

\medskip
\medskip

\setcounter{equation} {4}

\noindent {\bf Theorem 3.1} {\it Let $\vec B$ be a Bell function for which $LB\le TB < NSB$ holds, and let $\vec{PR}_1$ denote the PR box for which $\vec B\cdot \vec{PR}_1=NSB$. Let $TB^*\in [TB,NSB)$ and define $\mathcal Q_T= \{\vec P \in \mathcal{NS}|\vec B \cdot \vec P \le TB^*\}$. Then the set of extreme points of $\mathcal Q_T$ is equal to the set $\mathcal E$ defined as follows: all of the local deterministic distributions, all of the PR boxes except $\vec {PR}_1$, and all behaviors of the form
\begin{equation}
\vec E_i = \lambda_i\vec{PR}_1 + (1-\lambda_i)\vec{L}_i 
\end{equation}
where $\vec L_i$ is one of the eight local distributions saturating the version of the CHSH inequality maximally violated by $\vec{PR}_1$, and $\lambda_i=(TB^*-B_i)/(NSB-B_i)$ with $B_i = \vec B \cdot \vec L_i$.}

\medskip
\medskip

\setcounter{equation} {12}

\noindent \emph{Proof}. First note that the relation $B_i \le LB\le TB^* < NSB$ ensures that $0\le\lambda_i < 1$ holds. Thus \eqref{e:newextreme} defines a convex combination of the behaviors ${PR}_1$ and $\vec{L}_i$, so the $\vec E_i$ are valid behaviors in $\mathcal {NS}$. Furthermore, one can check directly that $\vec B \cdot \vec E_i\le TB^*$ holds for any $i$ (indeed, $\vec B \cdot \vec E_i= TB^*$) so the $\vec E_i$ are contained in $\mathcal Q_T$. As for the other elements of $\mathcal E$, these are all local deterministic distributions or PR boxes other than $\vec PR _1$, which all satisfy $\vec B \cdot \vec P\le LB\le TB^*$. Thus $\mathcal E\subseteq Q_T$. 

Now we demonstrate that $\mathcal E$ satisfies the statement of Definition \ref{d:extreme}. The first step is to show that every element of $\mathcal Q_T$ is in the convex hull of $\mathcal E$. To do this, note that the convex hull of $\mathcal E$ includes $\mathcal L$ as well as every convex combination of local deterministic distributions and PR boxes other than $\vec{PR}_1$. By the remarks following Theorem 2.2 of \cite{bierhorst:2016}, any element of $\mathcal{NS}$ that does not fall into one of these categories can be expressed as a convex combination of the following form:
\begin{equation}\label{e:orig}
p_{PR}\vec{PR}_1+\sum_{i=1}^8p_i\vec{L}_i.
\end{equation}
The $p$ coefficients are nonnegative and satisfy $p_{PR}+\sum_{i=1}^8p_i=1$. To show $\mathcal Q_T \subseteq \textnormal{Conv}(\mathcal E)$, it will thus suffice to show that any behavior $\vec P$ of the form \eqref{e:orig} that obeys $\vec B\cdot \vec{P}\le TB^*$ can be expressed as convex combination of the $\vec E_i$ behaviors defined in \eqref{e:newextreme} and the local distributions $\vec L_i$. 

Our strategy for this is to first re-write expression \eqref{e:orig} as
\begin{equation}\label{e:firststep}
\left[p_{PR}\vec{PR}_1+p_1\vec L_1 \right] + \sum_{i=2}^8p_i\vec{L}_i
=\left[p_{PR}'\vec{PR}_1+ p_1'\vec L_1+p_{E_1}\vec{E_1}\right] + \sum_{i=2}^8p_i\vec{L}_i
\end{equation}
where the new set of $p$ coefficients on the right side of \eqref{e:orig} are still nonnegative and sum to one, and the new coefficient $p_{PR}'$ for $\vec{PR}_1$ is smaller than $p_{PR}$. This process can be re-applied to $p_{PR}'\vec{PR}_1+p_2\vec{L}_2$ to further lessen the coefficient $p_{PR}'$, and the process is repeated while cycling from $\vec L_1$ through $\vec L_8$, until the entire weight of $PR_1$ is replaced with weight on the $\vec E$ vertices, leaving a convex combination solely of elements of $\mathcal E$.

To demonstrate that it is possible to execute the step displayed in \eqref{e:firststep}, we divide the problem into cases. First, let us suppose that $TB^*-B_1 \ne 0$ and that $p_1 \ge [(NSB-TB^*)/(TB^*-B_1)]p_{PR}$. Then the following equality holds:
\begin{equation}\label{e:swap1}
p_{PR}\vec{PR}_1+p_1\vec{L}_1 = p_{PR}\left(\frac{NSB-B_1}{TB^*-B_1}\right)\times \vec E_1 + \left[p_1-\left(\frac{NSB-TB^*}{TB^*-B_1}\right)p_{PR}\right]\times  \vec L_1.
\end{equation}
Our case assumptions assure that the coefficients of $\vec E_1$ and $\vec L_1$ are nonnegative, and furthermore the sum of the coefficients of $\vec E_1$ and $\vec L_1$ is equal to $p_{PR}+p_1$. Thus \eqref{e:swap1} can replace the $\vec{PR}$ and $\vec L_1$ terms in \eqref{e:orig} to obtain a well-defined convex combination of $\vec E_1$ and $\vec{L}_1,...,\vec{L}_8$ yielding the same behavior. In this case, containment in $\textnormal{Conv}(\mathcal E)$ is demonstrated and further iterations of the procedure are unnecessary.

Now let us suppose that either $TB^*-B_1 = 0$, or $TB^*-B_1 > 0$ and $p_1 < [(NSB-TB^*)/(TB^*-B_1)]p_{PR}$. In either case the following equality holds:
\begin{equation}\label{e:swap2}
p_{PR}\vec{PR}_1+p_1\vec{L}_1 = p_1\left(\frac{NSB-B_1}{NSB-TB^*}\right)\times \vec E_1 + \left[p_{PR}-\left(\frac{TB^*-B_1}{NSB-TB^*}\right)p_1\right]\times  \vec {PR}_1.
\end{equation}
Once again, the coefficients on the right side of the above equation are nonnnegative and sum to $p_{PR}+p_1$, and so \eqref{e:swap2} can replace the $\vec{PR}$ and $\vec L_1$ terms in \eqref{e:orig}. Now there will remain some weight on the PR box in the new convex combination, but it will be reduced (unless $p_1=0$ or $TB^*=B_1$, in which case it is unchanged), and the above process can be reapplied to $ \left[p_{PR}-\left(\frac{TB^*-B1}{NSB-TB^*}\right)p_1\right]\times \vec{PR}_1+p_2\vec{L}_2$. 

The key is that repeated applications of the \eqref{e:swap2}-type substitution to successive $\vec L_i$ terms must eventually terminate with a \eqref{e:swap1}-type substitution that eliminates the coefficient of $\vec {PR}_1$, prior to cycling through all eight of the $\vec L_i$. If this did not happen, the final expression after eight applications of \eqref{e:swap2}-type substitutions would be of the form $\sum_{i=1}^8\hat p_i\vec E_i + \hat p_{PR}\vec{PR}_1$ with $\hat p_{PR}>0$. But it is not possible for such an expression to be equivalent to \eqref{e:orig}, because $\vec B \cdot \vec E_i = TB^*$ for all $i$ and $\vec B \cdot \vec {PR}_1  > TB^*$, so the new expression would violate the bound $\vec B \cdot \vec P \le TB^*$, whereas the behavior in \eqref{e:orig} was assumed to obey it.

We have thus shown that $\mathcal Q_T \subseteq \textnormal{Conv}(\mathcal E)$. To complete the proof, we need to show the second part of Definition \ref{d:extreme} is satisfied; that is, no element of $\mathcal E$ can be expressed as a convex combination of other points in $\mathcal E$. We need only check this for new behaviors $\vec E_i$ defined in \eqref{e:newextreme}. Note first that if $TB^*-B_i=0$, then $\vec E_i=\vec L_1$, and $\vec L_i$ is already known to be extremal, so let us assume $TB^*-B_i>0$. In this case, $\vec E_i$ violates the version of the CHSH inequality that is maximially violated by $\vec{PR}_1$ and saturated by $\vec L_1$. This version of the CHSH inequality is not violated by any of the local deterministic distributions or PR boxes in $\mathcal E$, so any non-trivial convex combination of elements of $\mathcal E$ equaling $\vec E_i$ would require positive weight on at least one of the other behaviors defined by expression \eqref{e:newextreme}. But this is impossible: if we refer to Table 2 of \cite{bierhorst:2016}, we see that each $\vec E_i$ behavior looks like the one in Table \ref{t:eibehavior}.
\begin{table}[h]\caption{An example of a $\vec E_i$ behavior. From \eqref{e:newextreme}, $p_{PR}=(TB^*-B_i)/(NSB-B_i)$ and $p_i=(NSB-TB^*)/(NSB-B_i)$.}\label{t:eibehavior}
\centering
\begin{tabular}{ c|cccc }
 \multicolumn{1}{c}{}
  &  \multicolumn{1}{|c}{++}
 &  \multicolumn{1}{c}{+0}
 &  \multicolumn{1}{c}{0+} 
 &  \multicolumn{1}{c}{00}\\
  \cline{1-5}
$ab$ & $p_i+\frac{1}{2}p_{PR}$ & 0 & 0 & $\frac{1}{2}p_{PR}$\\
$ab'$ & $p_i+\frac{1}{2}p_{PR}$ & 0 & 0 & $\frac{1}{2}p_{PR}$\\
$a'b$ & $p_i+\frac{1}{2}p_{PR}$ & 0 & 0 & $\frac{1}{2}p_{PR}$\\
$a'b'$ & $p_i$ & $\frac{1}{2}p_{PR}$ & $\frac{1}{2}p_{PR}$ & 0\\
\end{tabular}
\end{table}
Importantly, there is a single location containing the entry $p_i$ and seven locations containing zero, all located outside the support of the PR box. Furthermore, each different $\vec E_i$ defined by \eqref{e:newextreme} will contain a positive $p_i$ entry in a {\it different} one of these eight cells outside the support of the PR box, and zeros in the rest. This feature implies that no convex combination equaling $\vec E_i$ can contain positive weight on any other $\vec E_j$ with $j\ne i$. $\hfill\Box$

\medskip
\medskip

\noindent Now we outline the proof that the extreme points of the polytope induced by $\vec B ^{\mathrm{CHSH}}$ and $\vec B ^\alpha$ are given by the expressions at the end of Section \ref{s:polytopes}. 

To demonstrate part 1 of Definition \ref{d:extreme}, consider an expression of the form \eqref{e:orig} that obeys both Tsirelson bounds. Then, analogously to \eqref{e:firststep}, replace a portion of the weight on the PR box with weight on one of the 16 behaviors of the form \eqref{e:doubleextreme}. This will be possible so long as there is positive weight on at least one $\vec L^{\mathrm{top}}_i$ behavior and at least one $\vec L^{\mathrm{bot}}_i$ behavior. Repeat this process with different choices of the 16 behaviors of the form \eqref{e:doubleextreme} until it is no longer possible -- either 1) all PR box weight has been converted, 2) there is no remaining weight on $\vec L^{\mathrm{top}}_i$ behaviors, or 3) there is no remaining weight on $\vec L^{\mathrm{bot}}_i$ behaviors. In case of (1), the process is complete; in case of (2), one can continue to replace PR box weight with $\vec E_i$-type behaviors as defined in \eqref{e:newextreme} with $\vec L^{\mathrm{bot}}_i$ behaviors (which saturate \eqref{e:tiltineq}, and strictly satisfy $\vec B ^{\mathrm{CHSH}}\cdot \vec P < 2\sqrt 2$); and in case of (3), one can instead continue to replace PR box weight with $\vec E_i$-type behaviors as defined in \eqref{e:newextreme} with $\vec L^{\mathrm{top}}_i$ behaviors (which, conversely, strictly satisfy \eqref{e:tiltineq} and saturate $\vec B ^{\mathrm{CHSH}}\cdot \vec P \le 2\sqrt 2$). The fact that the original behavior \eqref{e:orig} satisfies both Tsirelson inequalities ensures that in all cases, this process terminates with an equivalent expression consisting of a convex combination with weight solely on the new extreme points, and zero weight on the PR box. 

To demonstrate part 2 of Definition \ref{d:extreme}, note that the 24 new extreme points all violate the CHSH inequality, and the other extreme points do not, so any convex combination replicating one of the 24 new extreme points would have to contain weight on the other 23 new extreme points. However, it can be verified by inspection that this cannot occur by considering where these 24 extreme points contain zeros (recall Table \ref{t:eibehavior}) and which Tsirelson inequalities the extreme points saturate and/or strictly satisfy.

\bibliographystyle{unsrt}
\bibliography{metabib}
\end{document}